\documentclass[review]{elsarticle}
\usepackage{lineno,hyperref}
\modulolinenumbers[5]
\pdfoutput=1 
\usepackage[symbol]{footmisc}
\begin{document}

\begin{frontmatter}

\title{Generation of quasi continuous-wave electron beams in an L-band normal conducting pulsed RF injector for laboratory astrophysics experiments}

\author{Ye Chen\fnref{myfirstaddress,myfootnote1}}
\author{Gregor Loisch\fnref{myfirstaddress}}
\author{Matthias Gross\fnref{myfirstaddress}}
\author{Chun-Sung Jao\fnref{myfirstaddress}}
\author{\\Mikhail Krasilnikov\fnref{myfirstaddress}}
\author{Anne Oppelt\fnref{myfirstaddress}}
\author{Jens Osterhoff\fnref{mysecondaddress}}
\author{Martin Pohl\fnref{myfirstaddress,mythirdaddress}}
\author{\\Houjun Qian\fnref{myfirstaddress}}
\author{Frank Stephan\fnref{myfirstaddress}}
\author{Sergei Vafin\fnref{myfirstaddress,mythirdaddress}}

\address[myfirstaddress]{Deutsches Elektronen-Synchrotron DESY, Platanenallee 6, 15738 Zeuthen, Germany}
\address[mysecondaddress]{Deutsches Elektronen-Synchrotron DESY, Notkestraße 85, 22607 Hamburg, Germany}
\address[mythirdaddress]{Institute of Physics and Astronomy, University of Potsdam, Potsdam-Golm, Germany}

\fntext[myfootnote1]{ye.lining.chen@desy.de.}

\begin{abstract}
We report on an approach to produce quasi continuous-wave (cw) electron beams with an average beam current of milliamperes and a mean beam energy of a few MeV in a pulsed RF injector. Potential applications are in the planned laboratory astrophysics programs at DESY. The beam generation is based on field emission from a specially designed metallic field emitter. A quasi cw beam profile is formed over subsequent RF cycles at the resonance frequency of the gun cavity. This is realized by debunching in a cut disk structure accelerating cavity (booster) downstream of the gun. The peak and average beam currents can be tuned in beam dynamics simulations by adjusting operation conditions of the booster cavity. Optimization of the transverse beam size at specific positions (e.g., entrance of the plasma experiment) is performed by applying magnetic focusing fields provided by solenoids along the beam line. In this paper, the design of a microtip field emitter is introduced and characterized in electromagnetic field simulations in the gun cavity. A series of particle tracking simulations are conducted for multi-parametric optimization of the parameters of the produced quasi cw electron beams. The obtained results will be presented and discussed. In addition, measurements of the parasitic field emission (PFE) current (dark current) in the PITZ gun will be exemplarily shown to distinguish its order of magnitude from the produced beam current by the designed field emitter.
\end{abstract}


\end{frontmatter}


\section{Introduction}\label{Intro}

Laboratory astrophysics has drawn growing interest in the astrophysics community over the last decade \cite{King14, Compernolle15, Gekelman16, Warwick17, An17, Shukla17, Bell04, Nie10}. Alternatively to conventional methods of observation and numerical simulation, it is assumed to be an efficient way to improve our understanding of astrophysical processes in the environment of a scaled laboratory experiment. One interesting laboratory experiment considered at the Photo Injector Test facility at DESY in Zeuthen (PITZ) attempts to find the responsible mechanism for PeV-scale high energy cosmic-ray particles. As a promising candidate, the Bell’s instability \cite{Bell04,Nie10, CSHEDP}, a non-resonant instability driven by cosmic-ray current, is proposed to be the cause for those ultra-highly energetic particles by inducing turbulent amplification of the interstellar magnetic field in the upstream region of supernova remnant shocks. In order to study the Bell’s instability, a laboratory experiment using the PITZ accelerator was proposed \cite{CSHEDP, CSIFSA2017} based on the mechanism of electron beam driven magnetic instability growth in a matching plasma environment provided by the plasma cell of PITZ \cite{GDP2017}. Such an experiment sets specific requirements on quality parameters of the laboratory plasma and electron beams. Preliminary analysis in \cite{CSHEDP, CSIFSA2017} has shown that the electron beam parameters will be crucial for the occurrence of the Bell’s instability. This includes beam duration of milliseconds (ms), average beam current of milliamperes (mA), mean beam energy of a few MeV and a properly focused beam with a transverse size matched to the aperture of the plasma cell and the plasma density. A variety of advanced beam diagnostics is thus required to validate that the experimental conditions are fulfilled.

One solution to generate such electron beams in the PITZ gun is to use field emission (FE) from a metal based on the Fowler-Nordheim (FN) theory \cite{FNtheory}. In an RF (pulsed) regime this allows producing sub-ns electron bunches with peak currents of a few tens of mA during one period of the resonance frequency of the gun cavity (see details in Section \ref{FEBeam}). In comparison to the operation in the DC regime, the advantage of the RF gun performance thus remains in terms of higher beam energy \cite{FS2016}. However, gaps exist in between electron bunches generated from neighboring RF cycles preventing overall formation of a cw electron beam. Thus, the normal conducting 14-Cell L-band cut-disk accelerating structure downstream of the gun (booster) is used as a debunching cavity to lengthen the electron bunch within each RF period \cite{velobunching}. By properly tuning operation conditions of the booster a quasi cw electron beam can be formed by joining all individual bunches from neighboring RF cycles. It should be noted, that this idea could also be simply extended to similar solutions such as the use of a DC electron gun or a thermionic RF gun along with a booster cavity downstream, although technical issues may be different in each specific case. In this paper, we will focus only on the exemplary case of using an RF gun with field emission cathode.

The paper is organized as follows. In Section \ref{PITZoverview}, an overview of the PITZ facility is given. The RF gun of PITZ is described. In Section \ref{FEBeam}, measurements of the parasitic field emission current (i.e., dark current) in the PITZ gun are presented. A metallic field emitter placed on the backplane of the gun cavity is specially designed. It is used to enhance the local electric field gradient at a microtip for initiating pronounced field emission (FE). The FE based electron beams are then generated according to the Fowler-Nordheim theory. Using the modified electric field maps tracking simulations of field-emitted electron bunches are performed up to the entrance of the plasma cell in Section \ref{GandO}. The quasi cw beams are optimized by tuning operation conditions of the booster and a set of solenoids for fulfilling the experimental conditions as stated above. A summary and an outlook are given in Section \ref{Conclusion}. 

\section{Facility Overview and the RF Gun of PITZ}\label{PITZoverview}
The Photo Injector Test facility at DESY in Zeuthen (PITZ), was built to test, develop and experimentally optimize high brightness photoelectron sources for the operation of TESLA technology based Free Electron Lasers (FELs). The RF guns prepared at PITZ are in use at the Free electron LASer in Hamburg (FLASH) and the European X-ray Free Electron Laser (European XFEL). The PITZ LINAC consists of an L-band 1.6-Cell normal conducting (NC) RF gun, a pair of focusing solenoids, an NC 14-Cell booster cavity, and various advanced systems for electron beam diagnostics. The RF gun, the booster cavity and the plasma cell are located at 0, 2.7 and 6.0 meters (starting positions w.r.t. the cathode plane) in the beam line, respectively. The interested reader is referred to \cite{MKPRAB2012, FSPRAB2010} for a detailed description of the facility. Reports on plasma acceleration research activities at PITZ can be found in \cite{MG2016, GDP2016}.

\begin{figure}[!htbp]
\centering
\includegraphics[width=70mm]{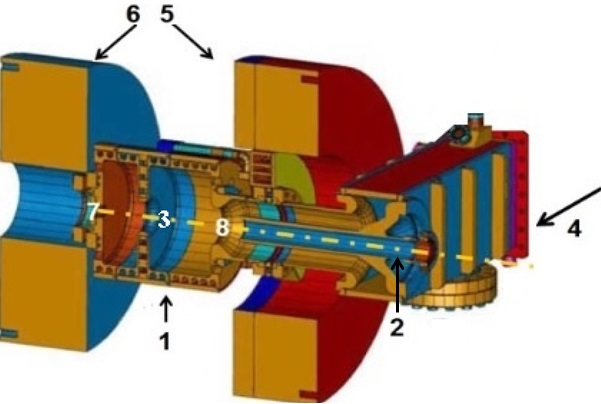}
\caption{Cut-view sketch of the PITZ RF gun and its surrounding solenoids:  1-gun cavity, 2-door-knob transition, 3-cavity axis, 4-RF feeding direction, 5-main solenoid, 6-bucking solenoid, 7-cathode and 8-end of coaxial line.}
\label{RFgun}
\end{figure}

The key component of the PITZ accelerator is an RF gun \cite{Klaus1997,Yves2017}, as illustrated in Fig. \ref{RFgun}. It is composed of a 1.3 GHz copper resonator operated in $\pi$ mode, coaxial RF power coupler, door-knob transition, input waveguide and supporting systems.

The PITZ gun can be operated with a high electric field gradient of about 60 MV/m on the cathode surface, resulting in a high peak RF power of about 6.5 MW in the cavity. It allows operation with long RF pulses of up to (650-1000) $\mu$s at a repetition rate of 10 Hz. This defines a high average RF power of more than 50 kW, dissipated in a rather short cavity length of about 20 cm. The gun has, meanwhile, shown a good RF amplitude stability of about $2\mathrm{e}{-4}$ and a phase stability of about 0.06 deg during its operation \cite{II2018}.

\section{Field Emitted Electron Beam in the Gun}\label{FEBeam}
Field emission (FE) from metals refers to quantum mechanical tunneling of conduction band electrons through the potential barrier at the surface of the metal. The FE-based electrons can be extracted when a GV/m-scale electric field gradient is locally formed at the field emitter (i.e., a needle cathode). This process is described by the Fowler-Nordheim (FN) equation \cite{FNtheory}. In an RF regime, the emission current is expressed as
\begin{eqnarray}
I=\frac{5.7\times10^{-12}\times10^{4.52\phi^{-0.5}}\times{A_e{{{E_s}}^{2.5}}}}{\displaystyle \phi^{1.75}}\times {\mathrm{exp}(-\frac{6.53\times10^{9}\times{\phi^{1.5}}}{{E_s}})},\label{eq:1}
\end{eqnarray}
and the RMS emission duration is derived as
\begin{eqnarray}
\sigma_t\approx\frac{1}{\omega}\sqrt[]{\frac{\beta{E_0}}{6.53\times10^{9}\times\phi^{1.5}}}.\label{eq:2}
\end{eqnarray}
The term $\textit{I}$ represents the peak FE current while $E_s$ characterizes the enhanced local electric field with $E_s$=$\beta{E}$=$\beta{E_0}{\mathrm{cos}{\omega t}}$, where $E$ stands for the macroscopic surface field oscillating with an angular frequency of $\omega$ and $E_0$ denotes the maximum amplitude of the field. The symbols $\phi$ and $A_e$ denote work function of the field emitter material and the effective emitting area, respectively. The field enhancement factor $\beta$ is defined here as the ratio of the enhanced local electric field $E_s$ to the original applied electric field $E$. The RMS emission duration per RF cycle $\sigma_t$ is a function of emitter material work function, field enhancement factor and maximum amplitude ($E_0$) of the RF electric field. Note in addition that all parameters are expressed in SI units.

Parasitic field emission (PFE) usually exists already in high field regions of an accelerating structure \cite{Han2005, Xiang2014}. This can for example refer to field emitted electrons born in the cathode area of the gun cavity, some of which may be captured by the accelerating fields and further transported downstream in the accelerator chain. Due to its complex nature of dynamics, the resulting PFE electron current, namely, dark current, is undesirable for accelerator operations. But, the dark current is usually several orders of magnitude lower compared to a controllable field emission beam current produced from a specially designed field emitter (see next subsections for details).  In the following, measurements of the dark current in the PITZ gun are shown first.

\subsection{Measurements of Dark Current in the PITZ Gun}
The dark current in the gun cavity is measured with a Faraday cup located at about 1.4 meters downstream from the cathode position. Characterization of the dark current is performed by varying the strength of the focusing magnetic field provided by the main solenoid in the gun section (see Fig. \ref{RFgun}).

\begin{figure}[!htbp]
  \centering
\includegraphics[width=70mm, height=30mm]{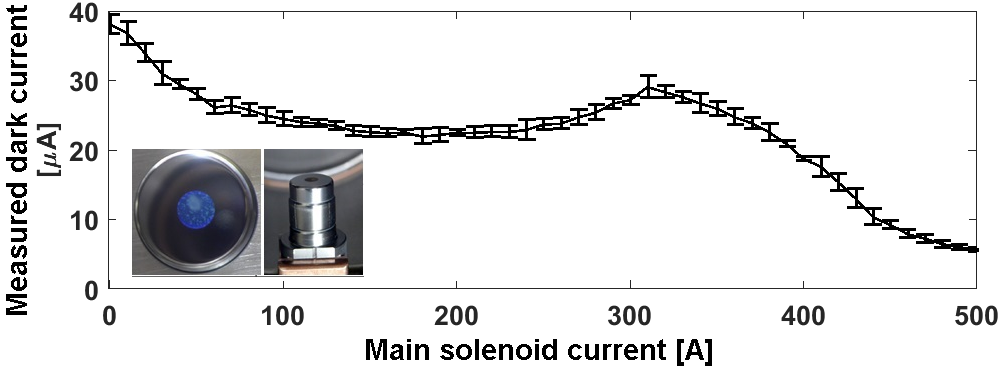}
\caption{The measured dark current as a function of the main solenoid current for a peak RF power of 6.4 MW in the gun. The inset pictures show the flat cathode and its plug used for these measurements.}\label{DCmeas}
  
   \vspace*{2mm}

\includegraphics[width=70mm, height=50mm]{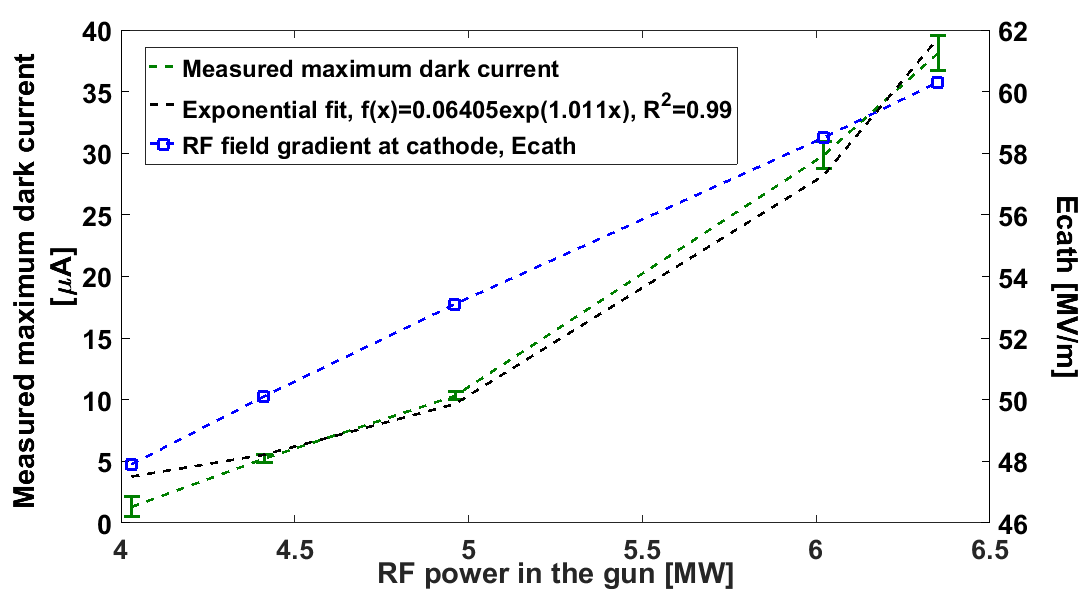}
\caption{The maximum dark current measured at different RF power levels in the gun for an RF pulse of 650 $\mu$s. Green curve: maximum measured dark current; Blue curve: maximum electric field gradients at the cathode position shown on the right axis.}\label{DCmeas2}

   \vspace*{2mm}
   
\includegraphics[width=65mm]{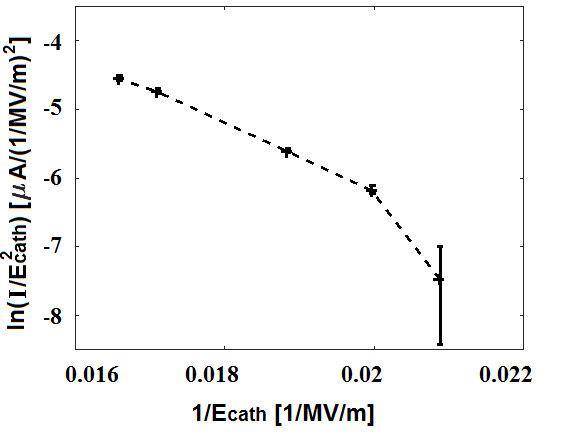}
\caption{The current-voltage characteristics described by the Fowler-Nordheim plot for the measurement data in Fig. \ref{DCmeas2}}\label{FNcurve}
\end{figure}

Figure \ref{DCmeas} shows the measured dark current in the gun cavity as a function of the main solenoid current when a flat cathode \cite{MKPRAB2012} is inserted in the cavity backplane. This is measured for an RF power of about 6.4 MW in the gun with a long RF macropulse of 650 $\mu$s. As shown, the maximum measured dark current is 38.1 $\mu$A. Figure \ref{DCmeas2} shows the maximum dark currents measured at 5 different RF power levels. The dark current shows exponentially increasing behavior as the RF power in the gun is increased (green curve). For a maximum peak RF power of about 6.4 MW, which corresponds to a peak electric field gradient of about 60 MV/m at the cathode position (blue curve on the right axis), the dark current is about 38 $\mu$A. The corresponding Fowler-Nordheim plot of the measurement data is shown in Fig. \ref{FNcurve}, which confirms the typical dependency of the field emission current on the electric field gradient. As shown in Figs. \ref{DCmeas}-\ref{FNcurve}, the dark current in the PITZ gun is roughly 2-3 orders of magnitude lower than the mA-scale field emission beam current produced by a specially designed field emitter (see next subsection). This is still true even for gun operation at a high RF power of 6.4 MW. This means, the parasitic field emission current would not make an obvious contribution to our beam current, and thus, can be neglected in the further design considerations.

\subsection{Design of a Microtip Field Emitter}
\begin{figure}[!htbp]
\centering
\includegraphics[width=80mm]{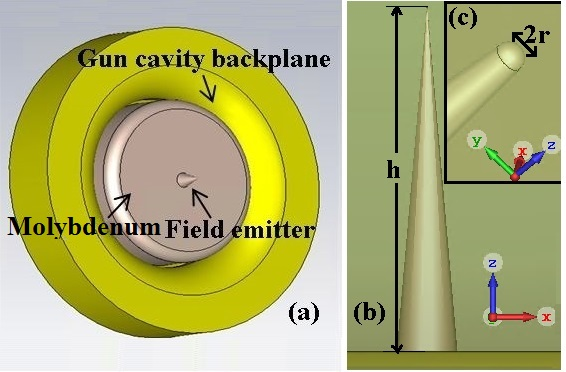}
\caption{Design of a single microtip field emitter. (a) Simulation model of the field emitter on the cavity backplane of the gun; (b) Cone type field emitter with a height of \textit{h} extending into the gun cavity; (c) Half sphere microtip with a radius of \textit{r}.}
\label{Emitter}
\end{figure}

In order to initiate pronounced and controllable field emission based beam current other than the background parasitic dark current, a dedicated microtip field emitter \cite{Roman2006,Roman2008} is specially designed as a field emission cathode. Molybdenum is chosen in our case as the material of the field emitter due to its high melting temperature and relatively low work function among metals. Most importantly, to obtain the required beam current, the geometry of the microtip field emitter needs to be optimized in the sense of sufficiently enhancing the originally applied electric field at the emitter position, and thus, providing a high local field enhancement factor to initiate strong field emission. A design of the field emitter is proposed and shown in Fig. \ref{Emitter}.

\begin{figure}
\centering
\includegraphics[width=70mm, height=55mm]{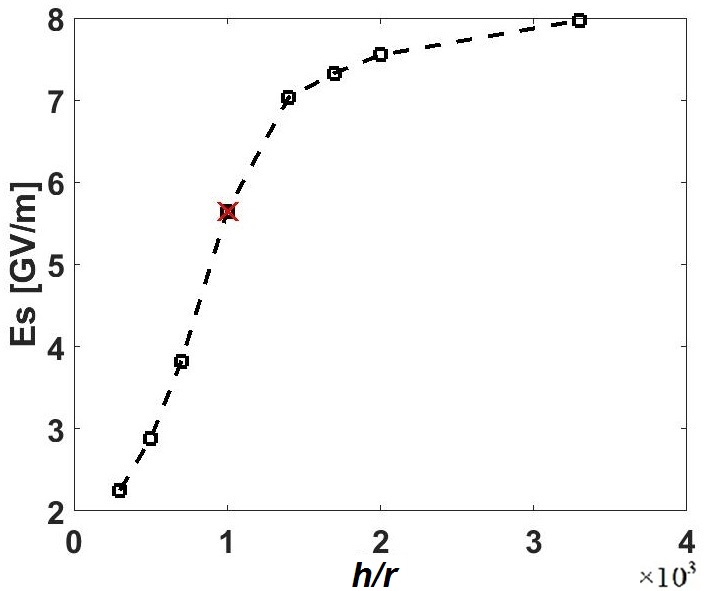}
\caption{Enhanced local electric field gradient at the position of the field emitter microtip as a function of \textit{h}/\textit{r}.}
\label{Es}
\end{figure}

As shown in Fig. \ref{Emitter}, a cone type field emitter made of molybdenum sits on the cavity backplane of the RF gun. A micrometer-scale half sphere emitter tip is incorporated on the top of the whole local field forming structure. The geometry optimization of the field emitter is carried out in terms of the ratio of the emitter height $\textit{h}$ to the microtip radius $\textit{r}$. In principle, longer emitter height extending into the gun cavity together with smaller radius of the microtip defines stronger local field enhancement effect. But, due to an increasing machining uncertainty with decreasing the fabrication scale of the microstructure, we take one micrometer as the radius of the microtip field emitter in later design considerations to keep the robustness of the design (see Fig. \ref{Es}). 

\begin{figure}[!htbp]
  \centering
\includegraphics[width=80mm]{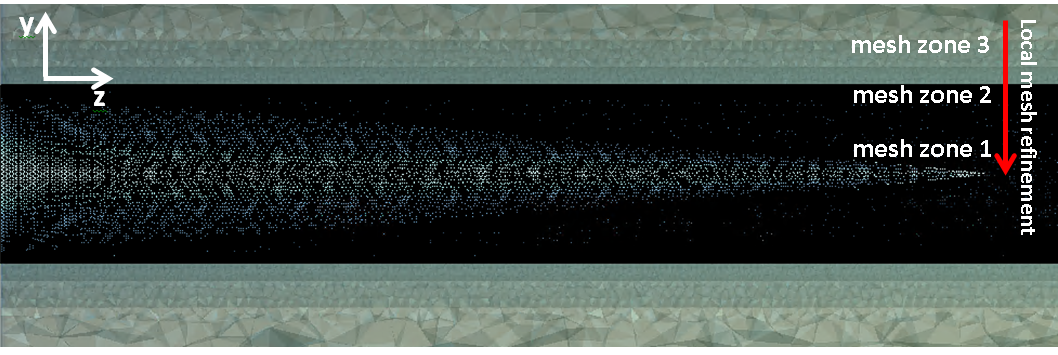}
\caption{Local meshing scheme around field emitter area. Gradual change of local meshing steps around the field emitter: mesh zone 1 as fine meshing, mesh zone 2 as medium meshing and mesh zone 3 as coarse meshing. Minimum mesh steps on the order of sub-micrometers and below is applied for best resolution.}\label{FEmeshing}
  
   \vspace*{2mm}

\includegraphics[width=80mm]{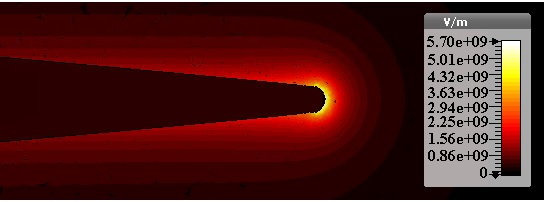}
\caption{Distribution of enhanced electric fields around the field emitter in the gun cavity. Note that the color map is scaled for better visualization of field enhancement effects.}\label{Es2}

   \vspace*{2mm}
   
\includegraphics[width=100mm]{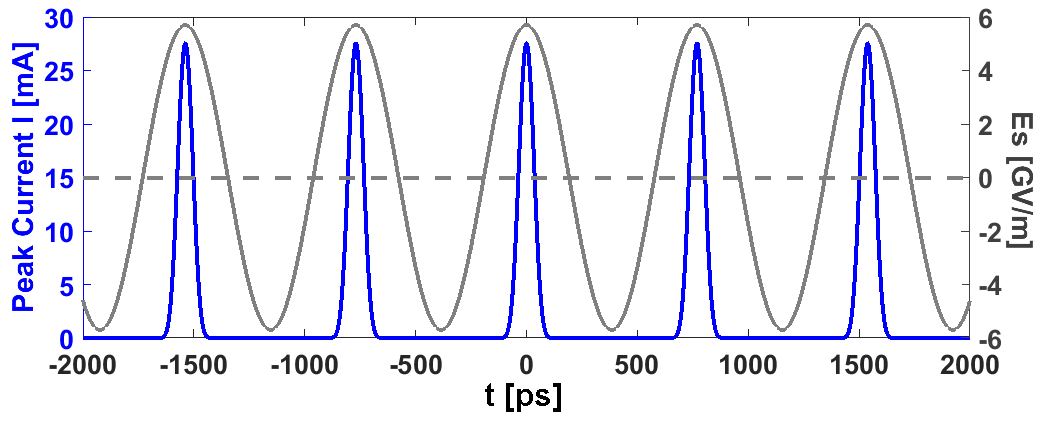}
\caption{FE-based electron beam over subsequent RF periods at 1.3 GHz and correspondingly the locally enhanced electric field gradient on the right axis.}\label{FNbeam}
\end{figure}

Figure \ref{Es} shows the enhanced electric field gradient at the field emitter that is calculated using the CST Microwave Studio software \cite{CST}. A typical multi-zone meshing scheme for the local mesh refinement is used to ensure numerical accuracy of the field calculation, as shown in Fig. \ref{FEmeshing}. In Fig. \ref{Es}, the enhanced electric field ($E_s$) is plotted as a function of the ratio of the emitter height ($\textit{h}$) to the microtip radius ($\textit{r}$). For these simulations, a maximum amplitude ($E_0$) of the electric field gradient of 60 MV/m is applied for a flat cathode in the RF gun. As shown, the enhanced local electric field is nonlinearly proportional to the geometrical parameter, $\textit{h}$/$\textit{r}$. For a robust design, a microtip radius of 1 $\mu$m and an emitter height of 1 mm can result in an enhanced electric field gradient of about 5.7 GV/m (marked by the red cross). This defines a local field enhancement factor ($\beta$) of roughly 95 at the microtip of the emitter. Figure \ref{Es2} shows the corresponding distribution of enhanced electric fields around the field emitter. In the following calculations, we choose $\beta$ = 95 exemplarily for demonstrating the generation of quasi cw electron beams. 

Applying $\beta$ = 95 and $\phi$ = 4.37 eV (work function of molybdenum) into Eqs. \ref{eq:1}-\ref{eq:2}, an FE-based electron beam can be generated according to the Fowler-Nordheim theory. As shown in Fig. \ref{FNbeam}, the temporal profile of the beam is similar to a Gaussian distribution (blue curve). In this case, the duration of the field emission per RF cycle is approximately 100 ps in full width half maximum. The blue curve shows the FE-based electron beam over 5 subsequent RF cycles at the resonance frequency of 1.3 GHz of the gun cavity. A peak current of about 27 mA is produced for a maximum enhanced electric field gradient of about 5.7 GV/m. The gray curve on the right axis shows the enhanced electric field gradient at the field emitter correspondingly.  

As shown, long gaps exist between the FE-based electron bunches of neighboring RF cycles. To form a cw beam at the position of the plasma cell where the experiment has to be conducted, a booster cavity downstream of the gun is used for lengthening the individual bunches within each RF cycle \cite{velobunching}.

\section{Generation and Optimization of Quasi cw Electron Beams}\label{GandO}
\begin{figure}[!htbp]
\centering
\includegraphics[width=100mm]{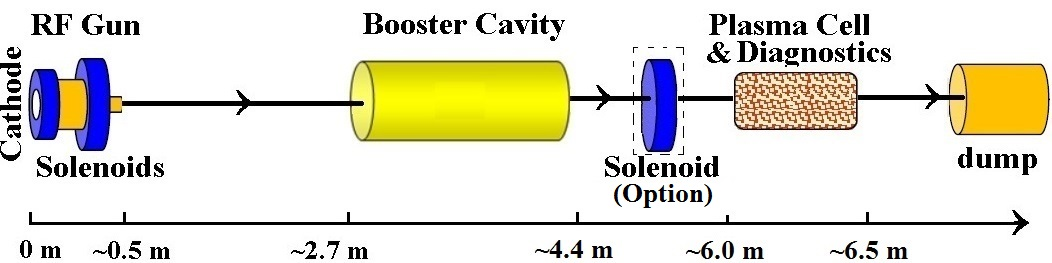}
\caption{Simulation setup for quasi cw electron beam generation and optimization. The diagram in the dashed frame indicates an option to insert an additional solenoid between the booster cavity and entrance of the plasma cell for further optimization of the transverse beam size. Note that this additional solenoid is currently not yet available but only for design considerations of beam dynamics.}
\label{Setup}
\end{figure}

Figure \ref{Setup} shows the main setup for the beam dynamics optimization of quasi cw electron beams in ASTRA \cite{ASTRA}. The FE-based electron beam, as shown in Fig. \ref{FNbeam}, is injected at the position of the field emitter in the gun. The beam is tracked through the gun cavity in the modified electric field map and focusing magnetic fields of the solenoids and further down through the booster cavity starting at about 2.7 m until the position of the plasma cell at about 6.0 m with respect to the position of the gun cavity backplane (0 m). A detailed description of the booster cavity at PITZ is given in \cite{MKPRAB2012}.

\begin{figure}[!htbp]
\centering
\includegraphics[width=80mm]{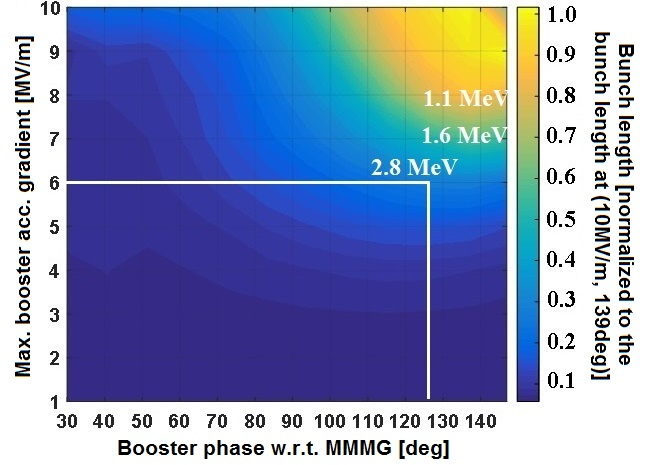}
\caption{Optimization of electron bunch length w.r.t. maximum booster accelerating gradient and operation phase.}
\label{2Dscan}
\end{figure}

\begin{figure}[!htbp]
\centering
\includegraphics[width=80mm]{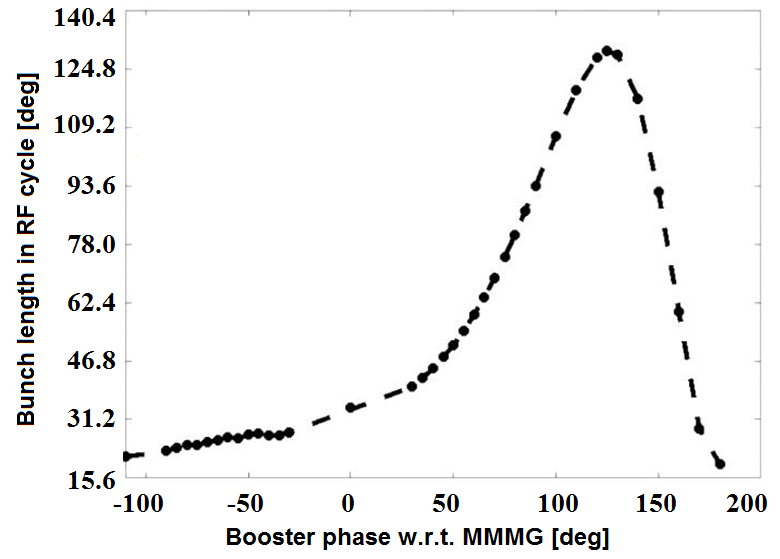}
\caption{Bunch length (RMS) in one RF cycle at the entrance of the plasma cell as a function of the booster phase for a maximum electric field gradient of 6 MV/m in the booster cavity. The optimum phase for the longest bunch length is at about 125 degrees w.r.t. the MMMG phase.}
\label{Zrms}
\end{figure}

In Fig. \ref{2Dscan}, operation conditions of the booster cavity are shown when optimizing for lengthening the electron bunches of subsequent RF periods. This is done by adjusting the maximum electric field gradient and operation phase of the booster cavity. The color code corresponds to the length of the electron bunch. The numbers in the plot indicate roughly corresponding beam energies at specific working points of the booster. As shown, the electron bunch is maximally de-bunched for higher booster gradients at operation phases far away from the maximum mean momentum gain (MMMG) phase. This trend is obvious at the upper right corner of Fig. \ref{2Dscan}. However, the trend shows a decreasing mean beam energy as the bunch length is increased. For the purpose of maintaining a beam energy above 2 MeV, we choose the working point at (6 MV/m, 125 deg) for the booster operation which results in a mean beam energy of 2.8 MeV. The projected bunch length in terms of the duration in one RF cycle versus booster phase is shown in Fig. \ref{Zrms} for the case with a booster field gradient of 6 MV/m.

\begin{figure}[!htbp]
\centering
\includegraphics[width=80mm]{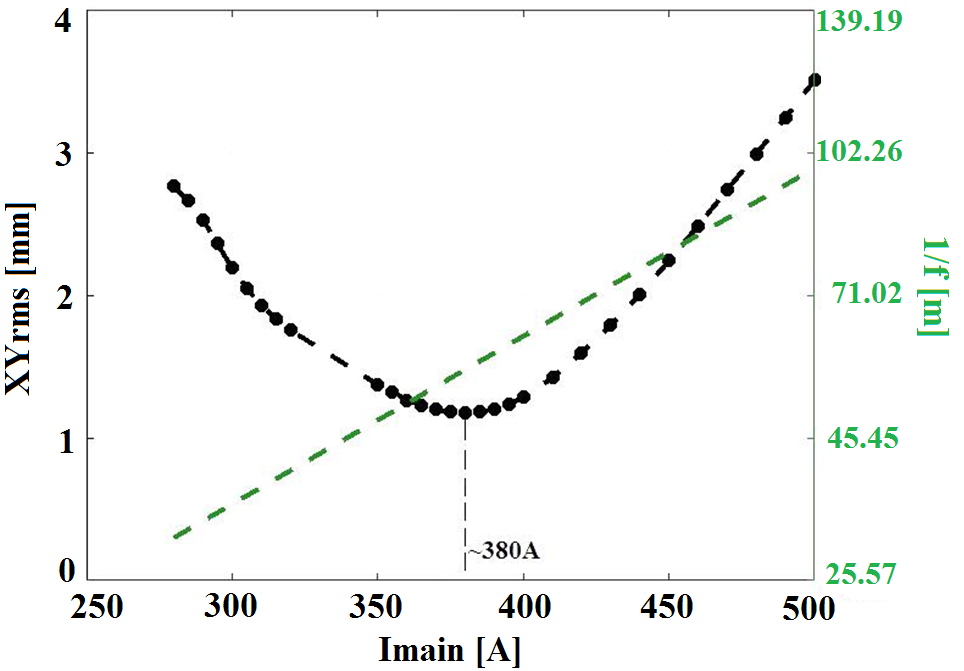}
\caption{Transverse beam size (RMS) at the entrance of the plasma cell versus the main solenoid current for a booster field gradient of 6 MV/m at 125 degrees w.r.t. the MMMG phase. The right axis represents the reciprocal of the focal length of the main solenoid.}
\label{Imain}
\end{figure}

Besides optimization of the debunching process with the booster cavity, the transverse focusing of the electron beam by solenoid fields also plays a crucial role. The transverse beam size has to be matched to the aperture of the plasma cell for efficient beam transport. It determines, furthermore, the electron beam density, which should be matched to the plasma density for initiating the Bell’s instability. In our case, the transverse beam size should be minimized to the smallest value possible at the entrance of the plasma cell. Figure \ref{Imain} shows the transverse beam size as a function of the main solenoid current while the corresponding effective focusing power (i.e., the reciprocal of the focal length) is shown on the right axis. As shown, the minimum beam size corresponds to a main solenoid current of about 380 A. The effective focal length ($\textit{f}$) is calculated using $f=\frac{8mE_k}{e^2\int_z \!{B_z}^2 \,\mathrm{d}z}$ based on \cite{IngoBz,BzBook}, where $\textit{e}$, $\textit{m}$ and $E_k$ denote fundamental electron charge, electron rest mass and kinetic energy, respectively. The term $B_z$ represents the magnetic field distribution of the main solenoid in the longitudinal ($\textit{z}$) direction.

\begin{figure}[!htbp]
  \centering
  \includegraphics[width=90mm, height=50mm]{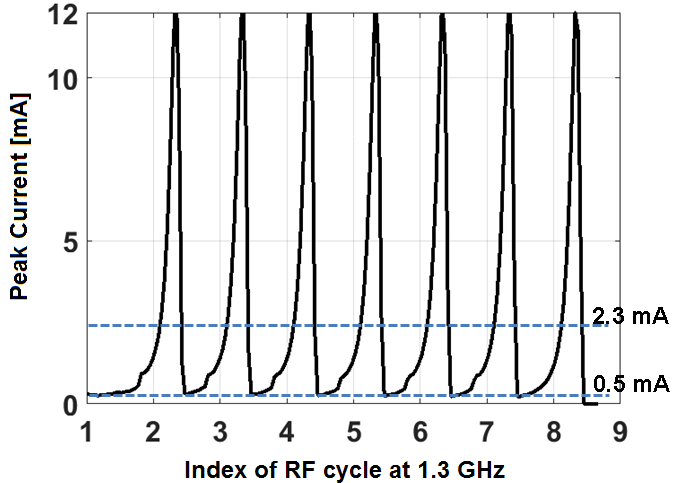}
  \caption{Quasi cw electron beam over 7 subsequent RF cycles at 1.3 GHz at the position of the plasma cell entrance. The base beam current $I_{base}$ is about 0.5 mA while the average beam current $I_{mean}$ is about 2.3 mA.}\label{CWbeam1}
  
   \vspace*{2mm}

  \includegraphics[width=90mm, height=35mm]{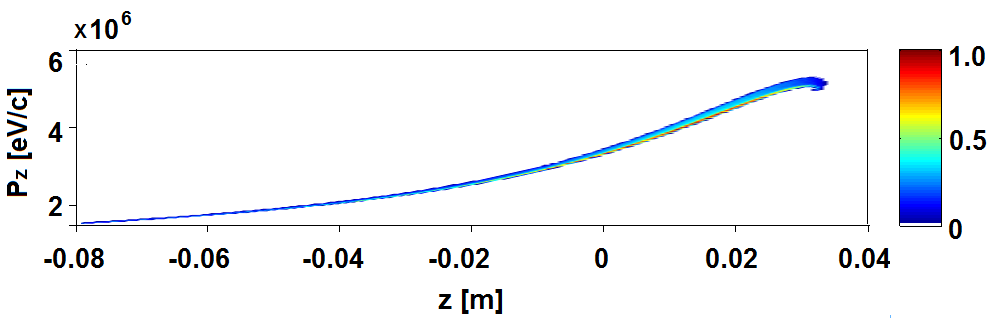}
  \caption{Longitudinal phase space distribution of the electron beam during one RF cycle.}\label{LPS1}

   \vspace*{2mm}
   
   \includegraphics[width=90mm, height=60mm]{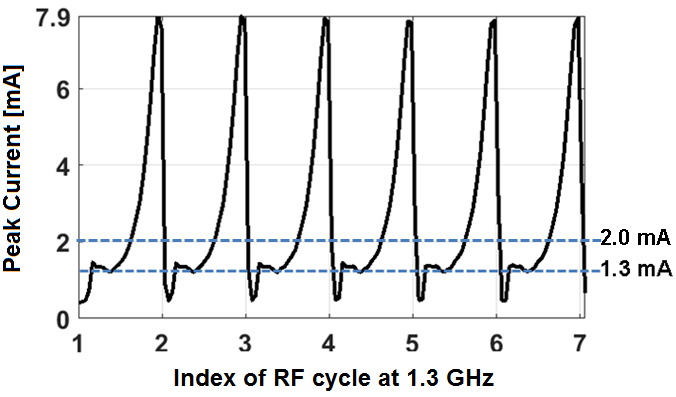}
   \caption{Quasi cw electron beam generation for the booster operation condition applied at (7 MV/m, 139deg). The base beam current is significantly increased to about 1.3 mA. A peak current of up to 7.9 mA, an average beam current of 2 mA and a mean kinetic beam energy of about 1.7 MeV can be expected.}\label{CWbeam2}
\end{figure}

In comparison to the injected train of electron bunches shown in Fig. \ref{FNbeam}, Fig. \ref{CWbeam1} shows the produced quasi cw electron beam over 7 RF cycles at $\lambda$ = 231 mm (i.e., 1.3 GHz) at the entrance of the plasma cell. For producing such a beam, the booster is operated with 6 MV/m at 125 degrees w.r.t. the MMMG phase. A main solenoid current of 380 A is applied. This set of operation parameters produces a peak beam current of 12 mA, an average beam current of about 2.3 mA, a transverse beam size of 1.1 mm (RMS) and a mean kinetic beam energy of about 2.8 MeV. As shown, with a base current of about 0.5 mA, the electron bunches from neighboring RF cycles together form a cw profile in the time domain. The longitudinal phase space (LPS) of the electron bunch during one RF cycle is shown in Fig. \ref{LPS1}. Note that the possible influences of the strong modulations in the current distribution of the produced quasi cw beam on inducing the Bell’s instability currently are still under investigations, and that the strength of the modulation can be modified by tuning the booster operation parameters.

It is also important to note, that the optimization process for the current profile of the electron beam, in terms of the modulation depth and the base current of the beam, essentially is a trade-off between a set of machine and beam quality parameters. For instance, varying operation condition of the booster cavity to (7 MV/m, 139 deg) would lead to a significant increase of the base beam current, as shown in Fig. \ref{CWbeam2}. In this case, a quasi cw beam with a base beam current of above 1 mA can be expected. However, the kinetic beam energy slightly drops down to about 1.7 MeV while an average beam current of 2 mA almost remains. Depending on specific conditions for the occurrence of the Bell’s instability in the laboratory, the electron beam parameters can be adjusted within a certain parametric range. This will be more precisely determined through ongoing plasma-astrophysics studies at DESY. 

\begin{figure}[!htbp]
\centering
\includegraphics[width=70mm]{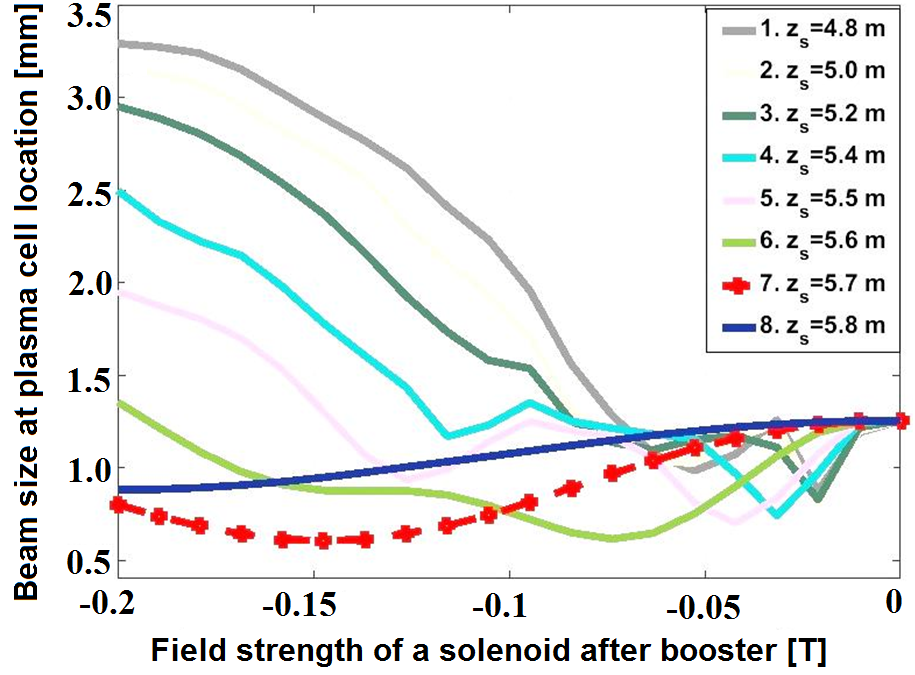}
\caption{Further optimization of the transverse beam size (RMS) by use of another solenoid between the booster and the plasma cell. The symbol \textit{$z_s$} in the figure legend represents the longitudinal position of the added solenoid.}
\label{Imain2}
\end{figure}

For further simulations, an additional solenoid located between the booster cavity and the plasma cell (see Fig. \ref{Setup}) is included for better matching the transverse beam size at the entrance of the plasma cell. This considers the position as well as the current of the added solenoid. Figure \ref{Imain2} shows the effect of the insertion of the solenoid at about 5.7 m with a magnetic field strength of about 150 mT. This may further reduce the transverse beam size to 0.55 mm (RMS) at the plasma cell entrance. However, due to a sharper focusing at the entrance, the beam size still grows by more than 10$\%$ over the full length of the plasma cell. Using the same conditions for booster and gun as in Fig. \ref{CWbeam1}, this improvement provides slightly better beam qualities: a peak current of 13.5 mA, an average beam current of about 2.33 mA and mean kinetic beam energy of about 3.05 MeV can be expected in this case.   

\section{Conclusion}\label{Conclusion}
A possibility of generating quasi cw electron beams has been studied in the L-band NC pulsed RF injector at PITZ for a planned laboratory astrophysics experiment at DESY. Field emission is used to produce an electron beam from a designed cone type field emitter with a half sphere microtip incorporated on the top of the cathode. A local field enhancement factor of 95 corresponding to an enhanced electric field gradient of 5.7 GV/m is simulated in the RF gun while a high electric field gradient of 60 MV/m is applied at the flat part of the cathode. Using FE-based electron beams, multi-parametric optimizations were performed for beam lengthening by the booster cavity and the beam transverse focusing by the solenoids was studied. This resulted in the following beam parameters: a peak beam current of 12 mA, an average current of about 2 mA, an RMS transverse beam size of about 1 mm and a mean kinetic beam energy of about 3 MeV. The base beam current over subsequent RF cycles at 1.3 GHz is shown to be above 0.5 mA confirming an uninterrupted beam temporal profile. The obtained results have shown the crucial beam astrophysics parameters that can be achieved for the laboratory experiment. A possible impact of the strong modulations in the longitudinal phase space of the quasi cw beam on inducing the Bell’s instability still needs to be investigated.  

\section{Acknowledgements}\label{Ack}
This research is partially supported by the DESY Strategy Fund.
\section*{References}

\bibliography{manuscript}

\end{document}